\documentclass[superscriptaddress,twocolumn,aps,prb,floatsfix]{revtex4}
\usepackage{graphicx}
\usepackage{amssymb,amsmath}
\usepackage{array}
\usepackage{dcolumn}

\begin{document}

\title{Leading interactions in the $\beta$-$S\!rV_6O_{15}$ compound}

\author{Marie-Liesse Doublet}
\affiliation{Laboratoire de Structure et Dynamique des Syst\`emes
    Mol\'eculaires et Solides, LSDSMS~/~UMR~5636,
    Universit\'e Montpellier 2, 
    Place Eug\`ene Bataillon, 
    F-34095 Montpellier Cedex 5, FRANCE}

\author{Marie-Bernadette Lepetit}
\affiliation{Laboratoire de Physique Quantique, 
    IRSAMC~/~UMR~5626,
    Universit\'e Paul Sabatier, 
    118 route de Narbonne, 
    F-31062 Toulouse Cedex 4, FRANCE}
\altaffiliation[On leave to~: ]{Laboratoire CRISMAT~/~UMR~6508, 
6 boulevard Maréchal Juin,
F-14050 Caen Cedex 4, FRANCE}

\date{\today}

\begin{abstract}
The present study shows that the electronic structure of the
$\beta$-$AV_6O_{15}$ family of compounds ($A = Sr, Ca, Na ...$) is
based on weakly interacting two-leg ladders, in contrast with the
zig-zag chain model one could expect from their crystal
structure. Spin dimer analysis, based on extended H\"uckel
tight-binding calculations, was performed to determine the structure
of the dominant transfer and magnetic interactions in the room
temperature $\beta$-$S\!rV_6O_{15}$ phase.  Two different two-legs
ladders, associated with different charge/spin orders are proposed to
describe these one-dimensional $\beta$-type systems.  The
antiferromagnetic ladders are packed in an 'IPN' geometry and
coupled to each other through weak antiferromagnetic
interactions. This arrangement of the dominant interactions explains
the otherwise surprising similarities of the optical conductivity and
Raman spectra for the one-dimensional $\beta$-type phases and the 
two-dimensional $\alpha$-type ones such as
the well-known $\alpha^\prime$-$N\!aV_2O_5$ system.
\end{abstract}

\maketitle

\section{Introduction}
Vanadium oxides are known since the fifties but they have attracted a
lot of attention in the recent years because of their exotic behavior.
Their remarkable properties are
due to the interplay between charge, spin and lattice degrees of freedom.
One of the most famous example is the $\alpha^\prime$-$N\!aV_2O_5$ phase
that undergoes a spin-Peierls transition~\cite{nav2-tr} coupled to a
spin ordering. This double transition has raised a large controversy
in the last five years before its nature and origin could be
elucidated, and before the apparent experimental contradictions could
be lifted. Indeed, the ordering associated with the transition was
supposed to be a charge ordering of the vanadium unpaired electron,
located on each rung of this two-legs ladder
system~\cite{nav2-tr2}. While vanadium NMR~\cite{nav2-rmn} and neutron
diffraction~\cite{nav2-neu} experiments exhibited a large charge
ordering at the transition, optical conductivity~\cite{nav2-co} and
resonant  X-ray diffraction~\cite{nav2-XR} did not show much charge
ordering. The controversy was lifted when ab-initio
calculations~\cite{vana3} showed that i) the bridging oxygens of the
ladder rungs have an open-shell character, ii) there are three and not
one magnetic electron per ladder rung and thus spin and charge
densities are not compelled to be equal, iii) the system presents a
large spin ordering (seen by spin sensitive experiments) associated
with a very weak charge ordering (as observed in charge sensitive
experiments).

Recently, superconductivity has been discovered in another family of
vanadium bronzes, renewing the interest in low-dimensional
vanadium oxides. The different phases of the $\beta$-$AV_6O_{15}$ ($A=
L\!i,\, N\!a,\,A\!g,\, C\!a,\, S\!r, C\!u$) family, also referred to
as $\beta$-$A_{0.33}V_2O_5$, exhibit an one-dimensional (1D) metallic
behavior at room temperature and undergo a metal to insulator phase
transition at $T_{MI}$~\cite{tmi-na}, associated with a charge
ordering.  Systems with monovalent cations ($A^+$) show a long-range
magnetic order at $T < T_{MI}$~\cite{longrangeMag}.  In systems with
divalent cations ($A^{2+}$), no sign of long-range magnetic order is
observed down to $2K$ and a spin gap appears in
$S\!rV_6O_{15}$~\cite{chord}.  The $\beta$-type phases present
crystal structures with a 1D arrangement along the $b$ direction
---~unlike the layered character of the $\alpha$-type phases. They
show six crystallographically independent vanadium atoms two-by-two
distributed over different cationic sites~: two $V_1$ atoms that form
zig-zag double chains composed of edge-sharing $VO_5$ square-based
pyramids, two $V_2$ atoms that form two-leg ladders composed of
$V_{2b}O_5$ square-based pyramid sharing a corner with $V_{2a}O_6$
distorted octahedron, and two $V_3$ atoms that form zig-zag
edge-sharing double chains, similar to the $V_1$'s, at first sights.

Some remarks should be done at this point.  \\ First, in the $\beta$-type systems, phases
with remarkable similitude of their structural arrangements,
whether they are doped by mono- or divalent cations, exhibit very
different magnetic properties.  Ueda {\it et al}~\cite{chord} suggested
that the differences in the magnetic properties are due to the
different nature of the electronically dominant subsystem. 
The electronically dominant subsystem would be the
zig-zag chains in the $\beta$-$A^+V_6O_{15}$ and ladders in
$\beta$-$A^{2+}V_6O_{15}$ compounds.
\\ Second, the $\alpha$ and $\beta$
compounds have very different structural arrangements ---~the former
is 2D while the latter is 1D~--- however their spectroscopic
properties such as Raman~\cite{cav6-rm} and optical conductivity
spectra~\cite{srv6-co} show remarkable similitudes.

A simple formal charge analysis gives $A^+ V^{(5-1/6)+}_6 O^{2-}_{15}$
for monovalent cations and $A^{2+} V^{(5-1/3)+}_6 O^{2-}_{15}$ for
divalent cations.  It comes a filling of either one or two $3d$
electrons for 6 vanadium atoms. This
(low) filling, as well as the strongly localized character of the
first row transition metal $3d$ orbitals, justify the use of a $t-J$
model on the vanadium sites to describe the low energy physics of
such vanadates.

\vspace{0.5cm}

The present paper thus aims at elucidating the structure of the
dominant electronic interactions in the $\beta$-type compounds within
the hypothesis of a $t-J$ model. For this purpose, spin dimer analysis
was performed using extended-H\"uckel tight-binding (EHTB) electronic
structure calculations. Since $\beta$-$AV_6O_{15}$ unit cells consist
in (at least) four formula units, $\it{i.e.}$ 88 atoms, this approach
offers a pertinent alternative to prohibitive ab initio calculations.
Furthermore, the EHTB method has been shown to provide reliable and
expedient means to study the relative strengths of both hopping and
spin exchange interactions in a wide variety of transition metal
oxides~\cite{whangbo}. Because strongly interacting spin exchange
paths of a magnetic solid are determined either by the overlap between
its magnetic orbitals for non-bridged interactions, or by the overlap
between its magnetic orbitals and the bridging ligand orbitals for
bridged interactions, a qualitative picture of both the dominant
magnetic interactions and their nature (antiferromagnetic versus
ferromagnetic) is reachable using EHTB, provided the knowledge of the
crystal structure. In the present work, calculations have been
performed using the crystallographic data reported for the
$\beta$-$S\!rV_6O_{15}$ compound~\cite{sellier}.

\section{Structure of the dominant interactions}

\begin{figure}[h]
{\includegraphics[width=6cm]{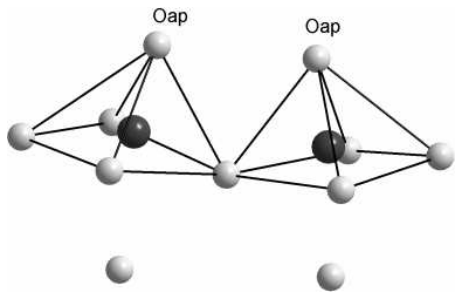}}
\caption{Local environment of vanadium atoms in the
$\beta$-$AV_6O_{15}$ phases. Vanadium and oxygen atoms correspond
to dark and light gray spheres, respectively. }
\label{f:pyr}
\end{figure}

In the $\beta$-type compounds, five of the six vanadium atoms are
surrounded by five nearest-neighbor (NN) oxygen atoms forming a
distorted squared pyramid ($V_{1a}$, $V_{1b}$, $V_{2b}$, $V_{3a}$ and
$V_{3b}$).  A sixth oxygen neighbor is found at a larger distance in
the position issued from a highly distorted $VO_6$ octahedron (see
figure~\ref{f:pyr}).  For the sixth vanadium atom $V_{2a}$, a less
marked pyramidal structure is observed, with two short, two medium and
two long $V-O$ bonds.
Each pyramid may actually be seen as a vanadyl $V\equiv O_{ap}$ cation
lying above a distorted square of $O^{2-}$ anions. Indeed, the
$V$--$O_{ap}$ distance is much smaller ($\sim 1.6$\AA) than the
$V$--$O$ distances involving either the oxygens contained in the
pyramid basal plane ($\sim 1.9$\AA--$2.0$\AA), or the sixth oxygen
($\sim 2.3$\AA). Let us note that the $V\equiv O_{ap}$ cations are not
rigorously perpendicular to the basal plane of the pyramids.  The
existence of an apical oxygen is crucial in these systems since the
short $V$--$O_{ap}$ distance allows a strong delocalization to occur
between these two atoms and a multiple vanadium--oxygen covalent bond
to take place. This phenomenon has already been observed in the
$\alpha^\prime$-$N\!a V_2 O_5$ compound, in which a triple
covalent/dative bond exists between the vanadium and its apical
oxygen. This bond is only weakly polarized~\cite{vana1} with an oxygen
charge of about $-0.5$. The experimental signature of such a strong
multiple bond in the Raman spectra is a sharp peak at
relatively high energy, corresponding to the bond stretching
mode. This peak occurs at $969\rm cm^{-1}$ for the
$\alpha^\prime$-$N\!aV_2O_5$~\cite{nav2-rm}, at $932\rm cm^{-1}$ and
$1002\rm cm^{-1}$ for the $C\!aV_2O_5$ and $M\!gV_2O_5$~\cite{v2-rm}
respectively. 
In the $\beta$-$C\!aV_6O_{15}$~\cite{cav6-rm}, it is seen at 
$978\rm cm^{-1}$, $952\rm cm^{-1}$ and $877\rm
cm^{-1}$ for the $V_3$, $V_1$ and $V_2$ vanadyl bonds.  Note that the softening of
the apical bond stretching mode for the $V_2$ atoms is due to a less
marked pyramidal character of its oxygen first neighbors.

\begin{figure}[h]
\includegraphics[width=8cm]{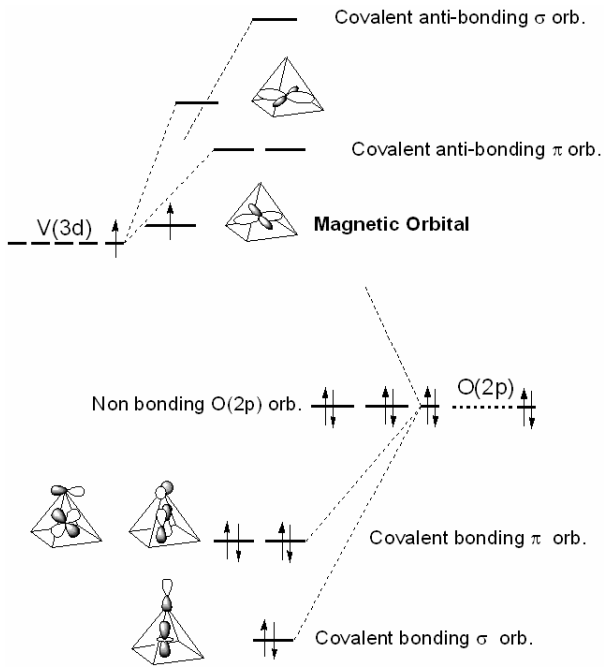}
\caption{
Qualitative orbital interaction between the vanadium $3d$
orbitals and the $p$ orbitals of the pyramid oxygens.
The apical oxygen lies on the $z$ axis, while the four 
oxygens of the pyramid basal plane lie on the $x$ and $y$ axes.}
\label{f:lf}
\end{figure}

Two consequences arise from the occurrence of such a $V$--$O_{ap}$
multiple bond. First the formal charge of the vanadium atom is much
smaller than what is usually assumed. It should be accounted as $5 -
\eta - q$ instead of $5 - \eta$ where $\eta$ is the number of magnetic
$3d$ electrons per vanadium atom and $q$ is the number of vanadium
electrons participating to the vanadyl bond. Second and much more
important, the vanadyl bond acts as a local quantification axis for
the vanadium atom.  As a consequence, the nature of the $3d$ magnetic
orbital can be deduced from the vanadyl bond orientation~: it is the
$d_{xy}$ orbital, when local axes are chosen so that $z$ is collinear
to the vanadyl bond and $x$ and $y$ point toward the basal plane first
oxygen neighbors (see figure~\ref{f:lf}).  Indeed, while the
$d_{z^2}$, $d_{xz}$ and $d_{yz}$ vanadium orbitals form one
$\sigma$ and two $\pi$ covalent/dative bonds with the apical oxygen,
the $d_{xy}$ and $d_{x^2-y^2}$ orbitals are split in agreement with
the crystal field. The $d_{x^2-y^2}$ orbital is more
destabilized than the $d_{xy}$ one due to its $\sigma$--type overlap
with the basal plane oxygen atoms which is larger than the $\pi$--type
overlap of the $d_{xy}$ orbital. Note that the structural difference
observed for the $V_{2a}$ local environment induces a lowering of one of 
the two anti-bonding $\pi$-orbitals close to the magnetic orbital energy 
and therefore a slightly different orientation of the $V_{2a}$ magnetic orbital. 

\begin{figure}[h]
\includegraphics[width=9cm]{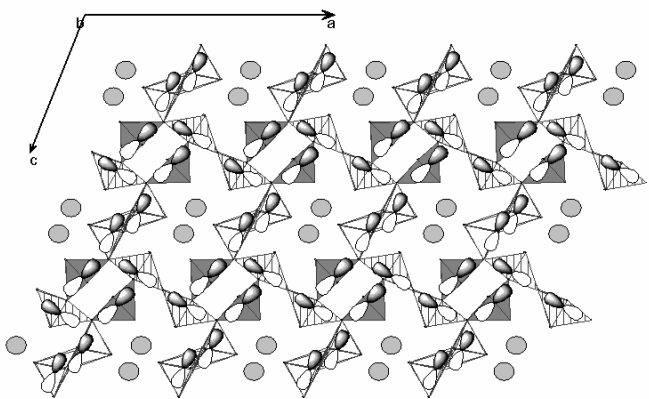}
\caption{Structural arrangement of the vanadium magnetic orbitals as
derived both from the vanadyl bond orientation on each vanadium atom
and from extended H\"uckel calculations. The crystal structure is
represented within the $(a,c)$ plane. The gray circles represent the
counter ions, the $V_2$ pyramids are hatched, the $V_3$ are
represented in white and the $V_1$ in gray.}
\label{f:omgac}
\end{figure}

Focusing now on the orientation of the vanadium magnetic orbitals in
the crystal structure, it becomes very simple to build the structure
of the magnetic orbitals once each vanadyl bond is located.
Figure~\ref{f:omgac} reports this structure for the
$\beta$-$S\!rV_6O_{15}$ system. The existence of two sets of vanadium
atoms can be clearly seen : on one hand the $V_1$ and $V_3$ centers
for which the magnetic orbitals are roughly in the $(c,b)$ plane and on
the other hand the $V_2$ centers for which the magnetic orbitals are
nearly orthogonal to the previous ones, in the approximate $(a+c,b)$
plane.

\begin{figure}[h]
\includegraphics[width=6cm]{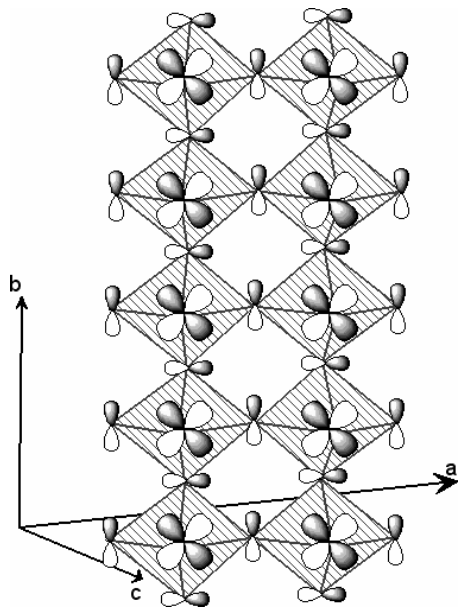}
\caption{
Vanadium magnetic orbitals of a $V_2$ ladder and $p$ orbitals
of the bridging oxygen atoms mediating the interactions between the
former.  Note that orbital signs do not have any meaning here. The
local $x$, $y$ and $z$ axes are defined as such~: the $z$ axis is in
the $V_{2b}$ vanadyl bond direction, the $x$ axis is in the ladder rungs
direction and the $y$ axis along the $b$ crystallographic direction. }
\label{f:omg2b}
\end{figure}

Let us now analyze separately each type of crystallographic chain or
ladder.  Figure~\ref{f:omg2b} reports the relative position of the
magnetic orbitals for the $V_2$ ladder.

\vspace{1cm}

Let us note that vanadyl bonds are nearly orthogonal to the figure,
pointing alternatively above and below the figure plane. The 
bridging-oxygen $p_x$ orbitals, along the ladder legs, and $p_y$
orbitals along the ladder rungs strongly overlap with the neighboring
$V_2$ magnetic orbitals. They will thus mediate both transfer and
magnetic interactions. This qualitative description is confirmed by
spin dimer EHTB calculations since the largest spin exchange paths
({\it i.e.} the largest effective transfer integrals) occur between
the $V_2$ magnetic centers, as stacked in the two-leg ladder shown in
figure~\ref{f:omg2b}. Due to significant local distortions of the
$V_2$-centered pyramids, transfer integrals are larger along the
ladder rungs ($\sim 0.31~eV$) than along the ladder legs ($0.17~eV$
and $0.21~eV$). Nevertheless, both effective integrals are large
enough to expect antiferromagnetic interactions between the $V_2$
atoms. Indeed, the effective transfer mediated through the appropriate
$p$ orbital of the bridging oxygen, can be expressed as
\begin{eqnarray} 
\label{eq:t}
 t &=&  \pm {t_{pd}^2 \over \Delta_1}
\end{eqnarray}
and thus the bridged super-exchange mechanism comes as 
\begin{eqnarray} 
\label{eq:j}
 J &=& - 4 { t_{pd}^4 \over \left(\Delta_1\right)^2 U_d}
       - 8 {t_{pd}^4 \over \left(\Delta_1\right)^2 \Delta_2}  \\
&=& -4 {t^2 \over U_d} - 8 {t^2 \over \Delta_2}
\end{eqnarray}
with $\; \; \Delta_1 = \delta - U_p + U_d - V_{pd}\; \;$ and $\; \;
\Delta_2 = 2\delta - U_p + 2U_d - 4V_{pd}\; \;$.  $\;\delta$ is the
orbital energy difference between the vanadium magnetic orbital and
the bridging oxygen $p$ orbital, $t_{pd}$ is the transfer integral
between them, $U_p$ and $U_d$ are the on-site Coulomb repulsions and
$V_{pd}$ is the bi-electronic repulsion between the V and O orbitals.

\begin{figure}[h] 
\includegraphics[width=7cm]{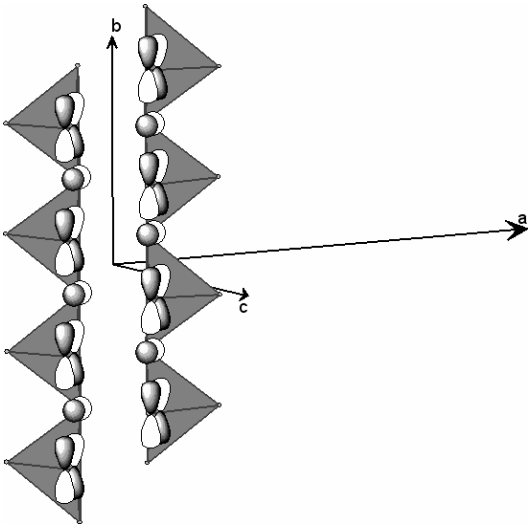}
\caption{Magnetic orbitals of the $V_1$ zig-zag chains and $p$
orbitals of the bridging oxygen atoms involved in the mediation of the
$V_1$ magnetic orbitals interaction. Note that the
orbitals represented here are atomic orbitals and that the relative phases
between them do not have any meaning. Local $x$, $y$ and $z$ axes
are defined as such~: $z$ is in the vanadyl bonds direction,
$y$ is along the $b$ crystallographic direction and $x$
is orthogonal to the formers. }
\label{f:omg1b}
\end{figure}

Figure~\ref{f:omg1b} reports  the relative position of the magnetic orbital
for the $V_1$ zig-zag chains.
One sees immediately that the hopping between the magnetic orbitals of
two nearest neighbors (NN) $V_1$ atoms is not bridged by any oxygen
orbital that could mediate the interaction, and is thus restricted to
its direct contribution. This direct contribution is itself expected
to be very small, due to both the large distance between adjacent
$V_1$ atoms ($\sim 3.3\AA$) and the $\delta$ character of
the overlap. EHTB calculations confirm these remarks since the
transfer integral between NN $V_1$ magnetic orbitals is found to be
negligible. As far as the magnetic exchange is concerned, no
super-exchange mechanism can take place between the NN $V_1$ atoms, for
the same reasons. Therefore, the effective magnetic interaction is
reduced to the direct exchange between the two vanadium atoms. It can
thus be predicted to be very weak and ferromagnetic. One can 
clearly consider that the NN $V_1$ atoms do not interact.  Let us now
examine the coupling between second neighbor $V_1$ atoms. One sees on
figure~\ref{f:omg1b} that the associated pyramids share a corner in
the $b$ direction and that the $p_x$ orbital of the bridging oxygen
presents a large overlap with the vanadium magnetic orbitals. It can
thus mediate efficiently the interaction between second neighbor $V_1$
atoms. EHTB calculations yield effective transfer integrals of
$0.17~eV$ and $0.15~eV$. Through-bridge super-exchange mechanism should take
place and the effective exchange between second neighbor $V_1$ atoms 
can thus be expected to by reasonably large and 
antiferromagnetic.

\begin{figure}[h]
\includegraphics[width=8cm]{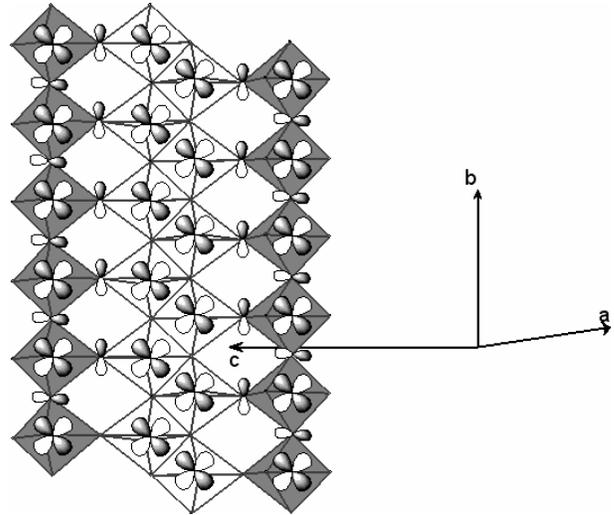} 
\caption{Magnetic orbitals of the $V_1$ and $V_3$ chains along the $b$
direction and $p$ orbitals of the bridging oxygen atoms.
Note that the orbitals represented here are atomic orbitals and that the
relative phases between them do not have any meaning. Pyramids
associated with the $V_1$ and $V_3$ atoms are represented in gray and white,
respectively.
Local $x$, $y$ and $z$ axes are defined as such~:  $z$  is in
the vanadyl bonds direction which is orthogonal to the figure plane,
$y$ is along the $b$ crystallographic direction and $x$
is orthogonal to the formers.}
\label{f:omg13b}
\end{figure}

Figure~\ref{f:omg13b} reports the relative position of the magnetic
orbitals of both the $V_1$ and $V_3$ chains.
The magnetic orbitals of the $V_3$ zig-zag chains are similar to the
magnetic orbitals of the $V_1$ atoms since the apical axes of both
sets of atoms are along the same direction. The $V_3$ pyramids share
an edge and the interaction between the $V_3$ magnetic orbitals is
bridged by two oxygen atoms with nearly $90^\circ$ $V_3$--O--$V_3$
angles. It is well known that in such $90^\circ$ arrangements, the
contribution of the bridging-oxygen $p$ orbitals is destructive and
that the interactions between the vanadium magnetic orbitals are
restricted to the direct transfer and exchange integrals. These
interactions are thus weak due to the large distance between two NN
vanadium atoms ($\sim 3.0\AA$). Indeed, EHTB calculations yield
effective transfers between nearest neighbor $V_3$ atoms of the order
of $\simeq 0.05~eV$ (see Appendix for the exact values of the four
crystallographically different $V_3$--$V_3$ transfer integrals). As
far as the effective exchange is concerned, one can expect it to be
both weak (of the order of a few tenth of meV or smaller) and
ferromagnetic ---~the direct exchange is always ferromagnetic in
nature. Let us now examine the second neighbor $V_3$--$V_3$
interactions. The associated pyramids share a corner along the $b$
direction and the $p_x$ orbital of the oxygen atom (not shown on
figure~\ref{f:omg13b}) can efficiently mediate the interactions
between two vanadium atoms. EHTB calculations yield transfer integrals
of $0.14eV$. Through-bridge super-exchange mechanism takes place via
the oxygens $p_x$ orbitals and the effective exchange between second
neighbors $V_3$ atoms should thus be reasonably large (hundreds of
meV) and antiferromagnetic.

It is clear from figure~\ref{f:omg13b} that there is another type of
large vanadium--vanadium interactions that is not considered in the
literature, that is the $V_1$--$V_3$ interaction. Indeed, the NN
$V_1$ and $V_3$ magnetic orbitals are bridged by a $p_y$ oxygen
orbital that mediates the interactions between them. EHTB
calculations confirm the present remarks with an evaluation of the
transfer integral of $0.20~eV$ and $0.23~eV$. Of course this $p_y$
oxygen orbital will mediate a super-exchange mechanism so that the
magnetic interaction can be expected to be large and
antiferromagnetic.

Summarizing the above results, one sees that the structure of the
dominant interactions, both transfer and magnetic, does not follow the
crystallographic zig-zag chains structure. Actually, the dominant
interactions are arranged in two types of two-leg ladders, namely the
$V_2$--$V_2$ ladders and the $V_1$--$V_3$ ladders which are
crystallographically orthogonal to each other. The average filling of
these ladders is of one electron for 3 sites in systems doped with
divalent cations and of one electron for 6 sites in systems doped with
monovalent cations. All magnetic interactions within these ladders
show an antiferromagnetic character.

Let us now analyze the  interactions between
these ladders. The coupling between two $V_1$--$V_3$ ladders could
go either through $V_1$--$V_1$ NN interactions which we have seen to
be totally negligible, both in terms of electron transfer and magnetic
interactions, or through the $V_3$--$V_3$ NN interactions. The latter
are somewhat larger than the NN $V_1$--$V_1$ interactions but still
quite weak and ferromagnetic in character. \\
Are there other types of inter-ladder interactions? Going back
to figure~\ref{f:omgac}, one sees that the $V_1$, $V_2$ and $V_3$
pyramids share a corner. 

Let us note the local axes in the following manner : $y$ for the b
direction, $x$ and $z$ in such a way that the $V_1$ and $V_3$ magnetic
orbitals are $d_{xy}$ and the $V_2$ magnetic orbitals are $d_{zy}$ in
nature.

The $p_y$ orbitals of the $V_1$--$V_2$--$V_3$ bridging oxygen mediates
the $V_1$--$V_3$ interaction.
One should notice that despite the nearly  $90^{\circ}$ $V_1$--$O$--$V_2$ and $V_3$--$O$--$V_2$ 
angles, the oxygen $p_y$ orbital can efficiently mediate the transfer interaction
between i) the  $V_1$ and $V_2$ magnetic orbitals and ii) the $V_3$ and $V_2$ 
magnetic orbitals. This is due to the fact that, unlike the classical case, 
the $V_n$ magnetic orbitals are
orthogonal to the ($V_1$,$O$,$V_2$) and ($V_3$,$O$,$V_2$) planes.
Indeed, EHTB calculations yield
$V_1$--$V_2$ and $V_3$--$V_2$ transfer integrals of the order of
$70~meV$ and $25~meV$. As far as the magnetic exchange is concerned, the
$p_y$ orbital is able to mediate super-exchange
mechanism. Thus the effective exchange between the $V_1$--$V_3$ and
$V_2$--$V_2$ ladders i) goes though the local $V_1$--$V_2$
and $V_3$--$V_2$ interactions, ii) can be expected to be weak but much
larger than the exchange between two $V_1$--$V_3$ ladders and iii) is 
antiferromagnetic in character.

\begin{figure}[h]
\includegraphics[width=8cm]{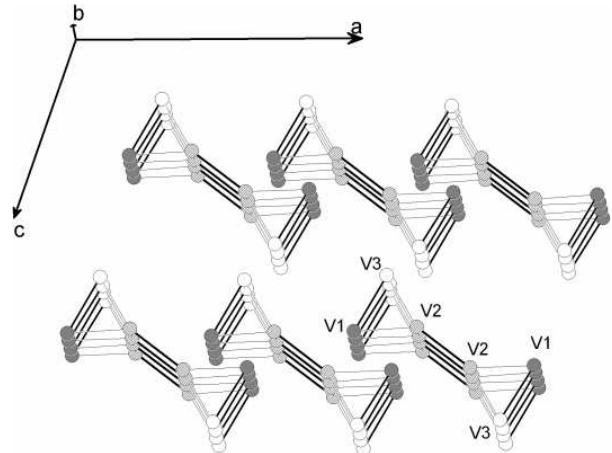}
\caption{Structure of the dominant interactions in the
$\beta$-$AV_6O_{15}$ compounds. Strong intra-ladder interactions are
in dark gray, while inter-ladder interaction are pictured in light gray.}
\label{f:IPN}
\end{figure}

In conclusion one can see the $\beta$-$AV_6O_{15}$ compounds as
composed of two types of orthogonal two-legs ladders packed in an
'IPN' geometry (see figure~\ref{f:IPN}) and coupled through
antiferromagnetic exchange interactions.

\section{On site energies}
In the $\beta$-$AV_6O_{15}$ compounds, the ladder rungs are built on
two crystallographically different vanadium atoms. This is obviously the case 
for the $V_1$--$V_3$ ladders and often for the $V_2$--$V_2$ ladders. 
For instance, even in the high
temperature phase, the $Sr$ and $Ca$ compounds present a
dimerization along the $b$ direction so that the $V_2$--$V_2$ ladder
rungs are composed of $V_{2a}$ and $V_{2b}$ crystallographically
different atoms. These crystallographic differences, which are system
and phase specific, are due to different chemical parameters such as
the size of the $A^{n+}$ cation, its location (in the $\beta$ or 
$\beta^\prime$ sites), or the electronic instabilities toward
spin-Peierls type of distortion. The consequence of these
characteristics is that the energies of the two vanadium magnetic orbitals 
in one ladder rung are different. EHTB calculations
show that these energy differences are not negligible, even
in the high temperature phases, since they can reach values as large
as $0.18~eV$ on the $V_2$--$V_2$ ladders. Surprisingly the $V_1$--$V_3$
ladders are much more symmetric with an orbital energy difference
between the $V_1$ and $V_3$ magnetic orbitals of $0.05~eV$ only.

The first consequence is that there should be a charge/spin order, even in
the high temperature phases. The second one is that the average
fillings, $\eta_{13}$ and $\eta_{22}$, of the two types of ladders have
no reason to be equal and that a charge transfer degree of
freedom between the two ladders should be taken into account.

\section{The low temperature phase}

Similar EHTB calculations as those detailed in the preceding sections
were done on the low temperature phase (LTPh). Structural data at 90K were
taken from reference~\cite{sellier}. In the low temperature phase the
unit cell is tripled along the ladder axis b. Sellier {\it et al} 
attributed this $3\times b$ super-cell to small displacements of the
$V_2$ atoms within the $V_2$ ladders rungs. 

As expected, the results in the LTPh exhibit on-site orbital energies
and transfer integral modulations compared to the high temperature
phase (HTPh), however the main conclusions remain unchanged. Indeed,
as in the HTPh, the dominant interactions are arranged in
antiferromagnetic ladders, and these ladders are then
antiferromagnetically coupled according to an IPN geometry. The main
difference between the two phases is that while in the HTPh the
transfer between IPNs is always very small, in the LTPh some of the
$V_3$--$V_3$ transfers are of the same order of magnitude as the
intra-IPN ones. Another point to be noticed is that the variation
range of the on-site orbital energies along the ladders is larger in
the LTPh than in the HTPh. These differences can thus be expected to
induce a greater electron localization in the ladder direction (in
agreement with the observed metal to insulator transition) and a
somewhat lesser 1D character in the LTPh.

\section{Discussion and conclusion} 
One of the troubling properties of the $\beta$-$AV_6O_{15}$ compounds is
that, while their crystallographic structure is very different from that of the
$\alpha^\prime$-$N\!aV_2O_5$ system, they present very similar features, both
in the optical conductivity and Raman spectra. The present study partially 
explains these similarities. Indeed, the electronic
structures of the two compounds are based on similar units, despite
their a-priori different crystallographic structures. They show (i)
square-pyramid environment of the vanadium atoms with magnetic
orbitals orthogonal to a very short vanadyl bond, and (ii) dominant
interactions of the magnetic centers arranged in two-leg ladders with
antiferromagnetic interactions.

The vanadyl multiple bond is responsible for the existence of a sharp
peak around $1000~{\rm cm^{-1}}$ in the Raman spectra of both $\beta$
and $\alpha$ compounds.  Other features related to the existence of
the pyramidal entities are present in the Raman spectra of both type
of compounds such as the bending mode between two pyramids around
$440~{\rm cm^{-1}}$. Finally, one retrieves in both compounds the broad
feature in the $550~{\rm cm^{-1}}$ to $700~{\rm cm^{-1}}$ range that
was attributed in the $\alpha^\prime$-$N\!aV_2O_5$ to the
electron-phonon coupling~\cite{nav2-rm2} responsible for the
spin-Peierls transition.

The optical conductivity spectra for both the $\beta$-$AV_6O_{15}$ and
the $\alpha^\prime$-$N\!aV_2O_5$ compounds also present strong
similitudes, such as the famous $1~eV$ peak. This peak was attributed
in the $\alpha^\prime$-$N\!aV_2O_5$ system to the first
doublet-doublet excitation energy on the ladder rung. It is also
present in the $\beta$-type compounds, which are ladder
systems, as well. In the $S\!rV_6O_{15}$, this peak is double ($0.85~eV$ and
$1.2~eV$~\cite{srv6-co}).  In the assumption that this peak can also be
attributed to the first doublet-doublet excitation energy of the
ladder rungs, the doubling would be in total agreement with the
electronic structure proposed in this work, namely two-leg ladders of
two different types.

\centerline{}

The occurrence of this peak in the $\beta$-type compounds raises several more
general questions on 1D vanadium oxides.\begin{itemize} 
\item The first one is
whether the existence of this $1~eV$ feature is the signature of a ladder
arrangement of the dominant interactions in the vanadium oxides. 
\item It has been shown in the $\alpha^\prime$-$N\!aV_2O_5$ compound
that the ladder rung should not be seen as supporting one electron
delocalized on the two vanadium magnetic orbitals, but rather three
magnetic electrons, since the bridging oxygen shows a strong open-shell
character and the local wave-function is  multiconfigurationnal~\cite{vana1, vana3}.
\\ The second question is thus whether this magnetic character of the
rung bridging oxygen is a general feature of the vanadium oxides with
a two-leg ladders electronic structure. 
\item The third question is whether the $1~eV$ peak is the signature of
this magnetic character of the rung oxygens.  One of us is actually
running ab-initio calculations in order to check these questions in
the $\beta$ compounds. The preliminary results confirm these hypotheses.
\end{itemize}
Finally let us remember that the sodium phase of the
$\beta$-$AV_6O_{15}$ family presents a super-conducing phase. Put into
perspective with the present results yielding a ladder structure for
the dominant interactions, one can wonder whether this compound could
be a realization of the predicted superconductivity in doped ladder
systems~\cite{EchSupra}.

\vspace{1.0cm}

{\bf Acknowledgments :} the authors thank the following research
groups for helpful discussions, namely E. Janod, Cl. Sellier, and
co-workers from the {\it Institut des mat\'eriaux Jean Rouxel},
V. Ta Phuoc from the {\it Laboratoire d'\'electrodynamique des
mat\'eriaux avanc\'es}, Y. Fagot-Revurat and co-workers from the {\it
Laboratoire de Physique des Mat\'eriaux}. We also thank the Groupement
de Recherche 2069 of the CNRS,{\it Oxydes \`a propri\'et\'es
remarquables}, for having made this collaboration possible.

\section*{Appendix}

\begin{table}[h]
\begin{tabular}{cdd|cdd}
\multicolumn{6}{c}{$V_1$--$V_3$ ladder} \\
\multicolumn{3}{c|}{Rungs} & \multicolumn{3}{c}{Legs} \\ 
Atoms &\multicolumn{1}{c}{$t$} & \multicolumn{1}{c|}{$\Delta \varepsilon $} & 
Atoms &\multicolumn{1}{c}{$t$} & \multicolumn{1}{c}{$\Delta \varepsilon $} \\
\hline
$V_{1b}$--$V_{3b}$ & -0.203 & 0.050 & $V_{1a}$--$V_{1b}$ & -0.173 & 0.00 \\ 
$V_{1a}$--$V_{3a}$ & -0.237 &0.113 & $V_{1a}$--$V_{1b+b}$ & -0.145 & 0.00 \\
&&& $V_{3a}$--$V_{3b}$ & -0.148 & -0.038 \\
&&& $V_{3a}$--$V_{3b+b}$ & -0.140 & -0.040 \\
\hline 
\hline
\multicolumn{6}{c}{$V_2$--$V_2$ ladder}  \\
\multicolumn{3}{c}{Rungs} & \multicolumn{3}{c}{Legs} \\
Atoms &\multicolumn{1}{c}{$t$} & \multicolumn{1}{c|}{$\Delta \varepsilon $} & 
Atoms &\multicolumn{1}{c}{$t$} & \multicolumn{1}{c}{$\Delta \varepsilon $}  \\
\hline
$V_{2a}$--$V_{2b}$ & -0.313 & -0.184 & $V_{2a}$--$V_{2b}$ & -0.147 & -0.173 \\
&&&$V_{2a}$--$V_{2b+b}$ & -0.135 & 0.215 \\
\hline \hline
\multicolumn{6}{c}{Inter-ladder Intra-IPN } \\[1ex]
Atoms &\multicolumn{1}{c}{$t$} & \multicolumn{1}{c|}{$\Delta \varepsilon $} &
Atoms &\multicolumn{1}{c}{$t$} & \multicolumn{1}{c}{$\Delta \varepsilon $} \\
$V_ {2a}$--$V_{1a}$ & 0.070 & 0.135 & $V_{2a}$--$V_{3a}$ & -0.024 & -0.022 \\
$V_{2b}$--$V_{1b}$ & 0.094 & -0.046 & $V_{2b}$--$V_{3b}$ & 0.054 & 0.093 \\
\hline \hline 
\multicolumn{6}{c}{Inter-ladder Inter-IPN } \\
Atoms &\multicolumn{1}{c}{$t$} & \multicolumn{1}{c|}{$\Delta \varepsilon $} &
Atoms &\multicolumn{1}{c}{$t$} & \multicolumn{1}{c}{$\Delta \varepsilon $} \\
$V_{3a}$--$V_{3a^\prime}$ & 0.023 & 0.002 & $V_{3a}$--$V_{3b^\prime}$ & -0.053 & -0.038 \\
$V_{3b}$--$V_{3b^\prime}$ & -0.059 & 0.006 & $V_{3b}$--$V_{3a-b}$ & -0.053 & -0.038 
\end{tabular}
\caption{EHTB values of the hopping and vanadium magnetic orbital energies (eV).}
\label{t:300}
\end{table}


\end{document}